\documentclass[cpp,a4paper,fleqn%
]{w-art}
\usepackage{times,cite,w-thm}
\usepackage[]{graphicx}
\usepackage{amsmath}
\usepackage{amssymb}
\usepackage{bm}

\newcommand{\aion}{a_\mathrm{i}}
\newcommand{\EF}{\epsilon_\mathrm{F}}
\newcommand{\mion}{m_\mathrm{i}}
\newcommand{\nion}{n_\mathrm{i}}
\newcommand{\Nion}{N_\mathrm{i}}
\newcommand{\kB}{k_B}

\newcommand{\omp}{\omega_\mathrm{p}}
\newcommand{\pF}{p_\mathrm{F}}
\newcommand{\TF}{T_\mathrm{F}}

\newcommand{\Tm}{T_\mathrm{m}}
\newcommand{\Tp}{T_\mathrm{p}}
\newcommand{\Prel}{P_\mathrm{r}}
\newcommand{\Tr}{T_\mathrm{r}}
\newcommand{\xr}{x_\mathrm{r}}
\newcommand{\gr}{\gamma_\mathrm{r}}
\newcommand{\tr}{\tau}

\newcommand{\dd}{\mathrm{d}}

\newcommand{\beq}{\begin{equation}}
\newcommand{\eeq}{\end{equation}}
\newcommand{\bea}{\begin{eqnarray}}
\newcommand{\eea}{\end{eqnarray}}
\newcommand{\req}[1]{Eq.\ (\ref{#1})}

\newcommand{\ApJ}[1]{{Astrophys.\ J.} \textbf{#1}}
\newcommand{\ApJS}[1]{{Astrophys.\ J. Suppl.\ Ser.} \textbf{#1}}
\newcommand{\AandA}[1]{{Astron.\ Astrophys.} \textbf{#1}}

\newcommand{\PL}[2]{{Phys.\ Lett. #1} \textbf{#2}}
\newcommand{\PR}[2]{Phys.\ Rev. #1 \textbf{#2}}
\newcommand{\PRL}[1]{{Phys.\ Rev.\ Lett.} \textbf{#1}}

\begin{document}

\edef\today{to be published in vol.\,50, issue 1 (Jan. 2010)}

\DOIsuffix{theDOIsuffix}
%%
%% issueinfo for header and copyright line
\Volume{49}
\Issue{12}
\Month{12}
\Year{2009}
%%
%%    First and last pagenumber of the article. If the option
%%    'autolastpage' is set (default) the second argument may be left empty.
\pagespan{1}{}
\Receiveddate{18 September 2009}
\Reviseddate{18 November 2009}
\Accepteddate{18 November 2009}
\Dateposted{}

%% \pretitle{Editor's Choice}

%% We have a short and a long form for the title. The short form
%% (optional argument) goes into the running head.

\title[Analytic equation of state of fully ionized plasmas]{Thermodynamic 
functions of dense plasmas: 
analytic approximations for astrophysical applications}

\author[A.~Y.\ Potekhin]{Alexander Y. Potekhin\footnote{Corresponding
     author: e-mail: {\sf palex@astro.ioffe.ru}}\inst{1,2}}
      \address[\inst{1}]{Ioffe Physical-Technical Institute,
     194021 St.\ Petersburg, Russia}
%%
%%    Information for the second author
\author[G.\ Chabrier]{Gilles Chabrier\inst{2}}
\address[\inst{2}]{Ecole Normale Sup\'erieure de Lyon,
     CRAL (UMR CNRS No.\ 5574),
     69364 Lyon Cedex 07, France}
\begin{abstract}
We briefly review analytic approximations of thermodynamic 
functions of fully ionized nonideal electron-ion plasmas,
applicable in a wide range of plasma
parameters, including the domains of nondegenerate and
degenerate, nonrelativistic and relativistic electrons, weakly
and strongly coupled Coulomb liquids, classical and quantum
Coulomb crystals.
We present improvements to previously published 
approximations. Our code for
calculation of thermodynamic functions
based on the reviewed approximations is made publicly available. 
\end{abstract}

\maketitle

%%  SECTION   ---------------------------
\section{Introduction}
\label{sect:intro}

In a previous work \cite{CP98,PC00}, we performed 
hypernetted chain (HNC)
calculations and proposed analytic formulae
for the equation of state (EOS) of electron-ion
plasmas (EIP). An alternative 
analytic approximation for the EOS of EIP was proposed in
\cite{SB96,SB00}. 
A comparison (e.g., \cite{CP98,SB00}) shows that 
the formulae in \cite{PC00} have a higher accuracy,
in particular
for the thermodynamic
contributions of ion-ion and ion-electron correlations
in the regime of moderate
Coulomb coupling.
Recently~\cite{PCR09,PCCDR09}, we studied classical ion mixtures
and proposed a
correction to the linear mixing rule.
In this paper we 
 review the analytic expressions
for all contributions to thermodynamic functions
and
introduce some
practical modifications to the previously published formulae.
The reviewed analytic
description of the thermodynamic
functions of Coulomb plasmas
is realized in a publicly available computer code.

%%  SECTION   ---------------------------
% \subsection{Plasma parameters}
% \label{sect:plasmpar}

Let $n_e$ and $\nion$ be the electron and ion number densities,
$A$ and $Z$ the ion mass and charge numbers, respectively.  The
electric neutrality implies $n_e = Z \nion$. In this paper we
neglect positrons (they can be described using the same formulae
as the electrons; see, e.g., Ref.~\cite{Blin}) and
free neutrons (see, e.g., Ref.~\cite{NSB1}), and consider mainly
plasmas containing a single type of ions (for the extension to
multicomponent mixtures, see \cite{PCR09,PCCDR09}).

The state of a free electron gas
is determined by the electron number density $n_e$ and
temperature $T$.
Instead of $n_e$ it is convenient to introduce
the dimensionless density parameter $r_s=a_e/a_0$,
where $a_0$ is the Bohr radius and
$a_e=(\frac43\pi n_e)^{-1/3}$.
The parameter $r_s$ can be easily evaluated from the relations
$
   r_s = 1.1723 \,n_{24}^{-1/3},
$
where $n_{24} \equiv n_e/10^{24}{\rm~cm}^{-3}$,
or
$
   r_s=(\rho_0/\rho)^{1/3},
$
where $\rho_0=2.6752\,(A/ Z)$ g~cm$^{-3}$.
The analogous density parameter for the
ion one-component plasma (OCP) is 
$R_S=\aion \mion (Ze)^2/\hbar^2 = 1822.89\, r_s  AZ^{7/3}$,
where $\mion$ is the ion mass
and $\aion\equiv(\frac43\pi \nion)^{-1/3}= a_e Z^{1/3}$ 
is the ion sphere radius.

At stellar densities it is convenient to use, instead of $r_s$,
the relativity parameter \cite{Salpeter61}
$
    \xr  = \pF / m_e c = 1.00884 \,
       \left( \rho_6  Z / A 
       \right)^{1/3}\!\! = 0.014005\,r_s^{-1},
$
where $ \pF = \hbar \,(3 \pi^2 n_e)^{1/3}
$
is the electron Fermi momentum and $\rho_6\equiv\rho/10^6$ g~cm$^{-3}$.
The Fermi kinetic
energy is
$\EF = c\,\sqrt{(m_e c)^2 + \pF^2}-m_e c^2 ,
$
and the Fermi temperature equals
$\TF \equiv \EF/\kB = \Tr  \, ( \gr - 1) ,$
where 
$\Tr \equiv {m_e c^2 / \kB } = 5.93\times
10^9~\mathrm{K}$,
$\gr \equiv \sqrt{1+ \xr^2}$,
and $\kB$ is the Boltzmann constant.
If $\xr\ll1$, then $\TF\approx 1.163\times10^6\, r_s^{-2}$~K.
The effects of special relativity are controlled by $\xr$
in degenerate plasmas ($T\ll \TF$) and by $\tr \equiv {T /\Tr}$
in nondegenerate plasmas ($T\gg \TF$).

The ions are nonrelativistic in most
applications. The strength of the Coulomb interaction of ions
is characterized by the Coulomb coupling parameter
$
   \Gamma =
        {(Z e)^2}/{\aion  \kB T} = \Gamma_e Z^{5/3},
$
where
$
    \Gamma_e \equiv { e^2 }/{ a_e \kB T}
   \approx {22.747}\,{ T_6^{-1}}
       \left(\rho_6 {  Z  }/{ A}
       \right)^{1/3}
$
and $T_6\equiv T/10^6$~K.

Thermal de Broglie wavelengths of the ions and electrons are
usually defined as
$
   \lambda_\mathrm{i} = \left({2\pi\hbar^2}/{ \mion \kB T}\right)^{1/2}
$
and
$
   \lambda_e = \left({2\pi\hbar^2}/{ m_e \kB T}\right)^{1/2}.
$
The quantum effects on ion motion are
important either at $\lambda_\mathrm{i}\gtrsim \aion$ or at 
$T \ll \Tp$, where
$
   \Tp \equiv (\hbar \omp/ \kB)
       \approx 7.832 \times 10^6\, (Z/A)\,\sqrt{\rho_6}\,\textrm{~K}
$
is the \emph{plasma temperature} determined by
the ion plasma frequency
$
    \omp = \left(  {4 \pi e^2 \,\nion}
     Z^2 /\mion \right)^{1/2}.
$
The corresponding dimensionless parameter is
$\eta \equiv \Tp/T$.

% \subsection{Free energy components}

Assuming commutativity of the kinetic
and potential energy operators and separation of the
traces of the electronic and ionic parts of
the Hamiltonian,
the total Helmholtz free energy $F$
can be conveniently written as
$
   F = 
   F_\mathrm{id}^{(i)} + F_\mathrm{id}^{(e)} 
   + F_{ee} + F_\mathrm{ii} + F_{ie}, 
$
where $F_\mathrm{id}^{(i)}$ and $F_\mathrm{id}^{(e)}$  denote the ideal free energy of ions
and electrons,  and the last three terms represent
an excess free energy arising from  the electron-electron,
ion-ion, and ion-electron interactions, respectively. 
This decomposition induces analogous decompositions
of pressure $P$, internal energy $U$, entropy $S$, the heat capacity
$C_V$, and
the pressure derivatives
$
   \chi_T=(\partial\ln P/\partial\ln T)_V
$ 
and
$
   \chi_\rho=-(\partial\ln P/\partial\ln V)_T.
$
Other second-order functions can be expressed through these
ones by Maxwell relations.

%%%%%%%%%%%%%%%%%%%%%%%%%%%%%%%%%%%%%%%%%%%%%%%%%%%%%%%%%%%
\section{Ideal part of the free energy}
\label{sect:id}
%%%%%%%%%%%%%%%%%%%%%%%%%%%%%%%%%%%%%%%%%%%%%%%%%%%%%%%%%%%

The free energy of a gas of
$\Nion=\nion V$ nonrelativistic classical ions is
$
   F_\mathrm{id}^{(i)} =
     \Nion \kB T \left[\ln(\nion\lambda_\mathrm{i}^3/M)-1 \right],
$
where $M$ is the spin multiplicity.
In the OCP, it can be written in terms of the
dimensionless plasma parameters
% (Sect.~\ref{sect:plasmpar})
 as
$
   {F_\mathrm{id}^{(i)}}={\Nion\kB T}
      \left[ 3 \ln \eta - 1.5 \ln \Gamma
        - 0.5 \ln(6/\pi)-\ln M -1 \right].
$

The free energy of the electron gas is given by
$
   F_\mathrm{id}^{(e)} =
   \mu_e N_e  - P_\mathrm{id}^{(e)}\,V,
$
where $\mu_e$ is the electron chemical potential.
The pressure and the number density are functions of
$\mu_e$ and $T$:
\beq
   P_\mathrm{id}^{(e)} =
 \frac{8}{3\sqrt\pi}\,\frac{\kB T }{ \lambda_e^3}
   \left[ I_{3/2}(\chi_e,\tr)
   + \frac{\tr}{ 2}I_{5/2}(\chi_e,\tr) \right],
\quad
   n_e =
           \frac{4}{\sqrt{\pi}\,\lambda_e^3}
   \left[ I_{1/2}(\chi_e,\tr)
   + \tr I_{3/2}(\chi_e,\tr) \right],
\label{n_e}
\eeq
where 
$\chi_e=\mu_e/\kB T$
(here, we do not include
the rest energy $m_e c^2$ in $\mu_e$) and
\beq
   I_\nu(\chi_e,\tau) \equiv \int_0^\infty
  \frac{ x^\nu\,(1+\tau x/2)^{1/2}
    }{ \exp(x-\chi_e)+1 }\,{\dd}x
\label{I_nu}
\eeq
is a Fermi-Dirac integral.
An analytic approximation for $\mu_e(n_e)$
has been derived in \cite{CP98}.

In Ref.~\cite{CP98} we calculated $I_\nu(\chi_e,\tau)$
using analytical approximations \cite{Blin}. These approximations
are piecewise: below, within, and above the interval
$0.6\leq\chi_e<14$. Their typical fractional accuracy  is a few
$\times10^{-4}$, the maximum error and discontinuities at the
boundaries reach $\sim0.2$\%. For the first and second
derivatives, the errors and discontinuities lie within 1.5\%. At
$\chi_e\geq14$ we use the Sommerfeld expansion
(e.g., \cite{Chandra})
\beq
     I_\nu(\chi_e,\tr) \approx
        \mathcal{I}_\nu^{(0)}(\tilde\mu)
          +\frac{\pi^2}{6}\tr^2 \mathcal{I}_\nu^{(2)}(\tilde\mu)
+ \frac{7\pi^4}{360}\tr^4 \mathcal{I}_\nu^{(4)}(\tilde\mu)
        + \ldots,
\label{Sommer}
\eeq
where we have defined $\tilde\mu=\chi_e\tr=\mu_e/m_e c^2$,
\beq
   \mathcal{I}_\nu^{(0)}(\epsilon)
   = \int_0^\epsilon \mathcal{I}_\nu^{(1)}(\epsilon')\,\dd \epsilon' 
   = \int_0^{x_0} \big(\sqrt{1+x^2}-1\big)^{\nu-1/2}
   \frac{x^2\,\dd x}{\sqrt{1+x^2}},
\quad
   \mathcal{I}_\nu^{(n+1)}(\tilde\mu)
   = \frac{\dd \mathcal{I}_\nu^{(n)}(\tilde\mu)}{\dd\tilde\mu} ,
\eeq
$\mathcal{I}_\nu^{(1)}(\epsilon)=\epsilon^\nu
\sqrt{2+\epsilon}$
and $x_0 \equiv \sqrt{\tilde\mu(2+\tilde\mu)}$.
In particular,
$$
\mathcal{I}_{1/2}^{(0)}(\tilde\mu)
 = [x_0\gamma_0-\ln(x_0+\gamma_0) ]/2,
\quad
\mathcal{I}_{3/2}^{(0)}(\tilde\mu)
 = x_0^3/3 - \mathcal{I}_{1/2}^{(0)}(\tilde\mu),
\quad
\mathcal{I}_{5/2}^{(0)}(\tilde\mu)
 = x_0^3\gamma_0/4 - 2 x_0^3/3
  + 1.25\,\mathcal{I}_{1/2}^{(0)}(\tilde\mu),
$$
where
$\gamma_0 \equiv \sqrt{1+x_0^2} = 1+\tilde\mu\,$
(note that, if $\tilde\mu=\tilde\epsilon$, where
$\tilde\epsilon\equiv\EF/m_e c^2$,
then $x_0=\xr$ and $\gamma_0=\gr$).

At small $\tilde\mu$, accuracy can be lost
because of numerical cancellations
of close terms of opposite signs 
in \req{Sommer} and in the respective partial derivatives. 
In this case we use the expansion
\beq
 I_\nu(\chi,\tau) = \! I^\mathrm{nr}_\nu(\chi)
       + \!\sum_{m=0}^\infty (-1)^m
        \frac{(2m-1)!!\,\tr^{m+1}}{4^{m+1} m!}
        I^\mathrm{nr}_{\nu+m+1}(\chi),
\quad
 I^\mathrm{nr}_\nu(\chi) = \int_0^\infty \frac{ x^\nu
   \,\dd x }{ e^{x-\chi}+1 },
\label{expan1}
\eeq
where $(2m-1)!!\equiv\prod_{k=1}^m (2k-1)$ should be replaced by
1 for $m=0$. At large $\chi$, the nonrelativistic Fermi integrals
$I^\mathrm{nr}_\nu(\chi)$ are calculated with the use of 
the Sommerfeld expansion.
Sufficiently smooth and accurate overall approximations for
functions $I_\nu(\chi_e,\tau)$ with $\nu=\frac12$, $\frac32$, and
$\frac 52$ are
provided by switching to Eq.~(\ref{expan1})
at $\chi_e\geq14$ and $\tilde\mu<0.1$, 
while retaining the terms
up to $m=3$.

The chemical potential at fixed $n_e$ is obtained
with fractional accuracy $\sim T^2/\TF^2\,$ by 
using Eqs.~(\ref{n_e}), (\ref{Sommer}), setting
$
\mathcal{I}_\nu^{(n)}(\tilde\mu)\approx
\mathcal{I}_\nu^{(n)}(\tilde\epsilon)
+\mathcal{I}_\nu^{(n+1)}(\tilde\epsilon)\,(\tilde\mu-\tilde\epsilon) ,
$
and dropping the terms that contain the product
$(\tilde\mu-\tilde\epsilon)\,\tr^2$.
Then  at $T\ll\TF$ 
\beq
\Delta\tilde\epsilon \equiv
   \tilde\epsilon-\tilde\mu \approx
    \frac{\pi^2\tr^2}{6\,\tilde\epsilon}\,
     \frac{1+2\xr^2}{\gr(1+\gr)} \, ,
\quad
 \Delta F\equiv F_\mathrm{id}^{(e)} - F_0^{(e)}
  \approx -\frac12\,TS_\mathrm{id}^{(e)}\approx
  -\Prel V\frac{\xr\gr\tr^2}{6},
\label{DeltaEF}
\eeq
where 
$
   F_0^{(e)} =
        ({ \Prel V}/{ 8\pi^2 })
         [ \xr\,(1+2\xr^2)\,\gr
      - \ln(\xr+\gr) ]-N_e m_e c^2 
$
is the zero-temperature limit \cite{Frenkel} (without
the rest energy $N_e m_e c^2$),
and
$\Prel \equiv m_e c^2\,(m_e c/\hbar)^3
    = 1.4218\times 10^{25}\mathrm{~dyn~cm}^{-2}$
is the relativistic unit of pressure.

Accordingly,
$
P_\mathrm{id}^{(e)} \approx
P_0^{(e)}+\Delta P,
$
where
$
   P_0^{(e)} =
      ({\Prel}/{8\pi^2})\,
          \left[\xr\left(\frac23\,\xr^2-1\right) \gr
             + \ln(\xr+\gr)\right],
$
and  $\Delta P = (\Prel/18)\,\tr^2 \xr\,(2+\xr^2)/\gr$. In this case,
$\chi_\rho^{(e)} \approx
\Prel\xr^5/9\pi^2\gr P_\mathrm{id}^{(e)},
$
$
\chi_T^{(e)} \approx
2\Delta P/P_\mathrm{id}^{(e)},
$
$
C_V^{(e)} \approx \pi^2 \kB
N_e\,\gr\tr/\xr^2.
$
In order to minimize numerical
jumps at the transition between the fit at 
$\chi_e<14$ and the Sommerfeld expansion at $\chi_e>14$,
 we multiply the expressions for $\Delta F$
by empirical correction factor
$(1+\Delta\tilde\epsilon/\tilde\epsilon)^{-1}$,
and those for $\chi_T^{(e)}$ and $C_V^{(e)}$ by
$[1+(4-2\xr/\gr)\,\Delta\tilde\epsilon/\tilde\epsilon]^{-1}$.

At $\xr<10^{-5}$ we replace $F_0^{(e)}$ and
$P_0^{(e)}$ by their nonrelativistic limits,
$F_0^{(e)}/V = \Prel \xr^5/10\pi^2$ and 
$P_0^{(e)} = \Prel\xr^5/15\pi^2\propto n_e^{5/3}$. In the
opposite case of $\xr\gg1$, one has $P_0^{(e)} =
\Prel\xr^4/12\pi^2\propto n_e^{4/3}$.

%%%%%%%%%%%%%%%%%%%%%%%%%%%%%%%%%%%%%%%%%%%%%%%%%%%%%%%%%%%
\section{Electron exchange and correlation}
\label{sect:ee}
%%%%%%%%%%%%%%%%%%%%%%%%%%%%%%%%%%%%%%%%%%%%%%%%%%%%%%%%%%%

Electron exchange-correlation effects were studied by
many authors. For the reasons explained in \cite{CP98}, we
adopt the fit to $F_{ee}$ presented in Ref.~\cite{IIT}. It is
valid at any densities and temperatures, provided that 
$\xr\ll1$.

In Ref.~\cite{PC00} we implemented an interpolation between the
nonrelativistic fit \cite{IIT} and approximation \cite{SB96,SB00}
valid for strongly degenerate
relativistic electrons. However, later on we found that such
an interpolation may cause an unphysical behavior of the heat
capacity in a certain density-temperature domain. Therefore we
have reverted to the original formula \cite{IIT},
taking into account that in applications, as long as the electrons
are relativistic, their exchange and correlation contributions are 
orders of magnitude smaller than the other contributions to the EOS of
EIP (see, e.g., \cite{YaSha,NSB1}).

%%%%%%%%%%%%%%%%%%%%%%%%%%%%%%%%%%%%%%%%%%%%%%%%%%%%%%%%%%%
\section{One-component plasma}
\label{sect:ii}
%%%%%%%%%%%%%%%%%%%%%%%%%%%%%%%%%%%%%%%%%%%%%%%%%%%%%%%%%%%

\begin{figure}
\begin{minipage}[t]{.48\linewidth}
    \includegraphics[width=\linewidth]{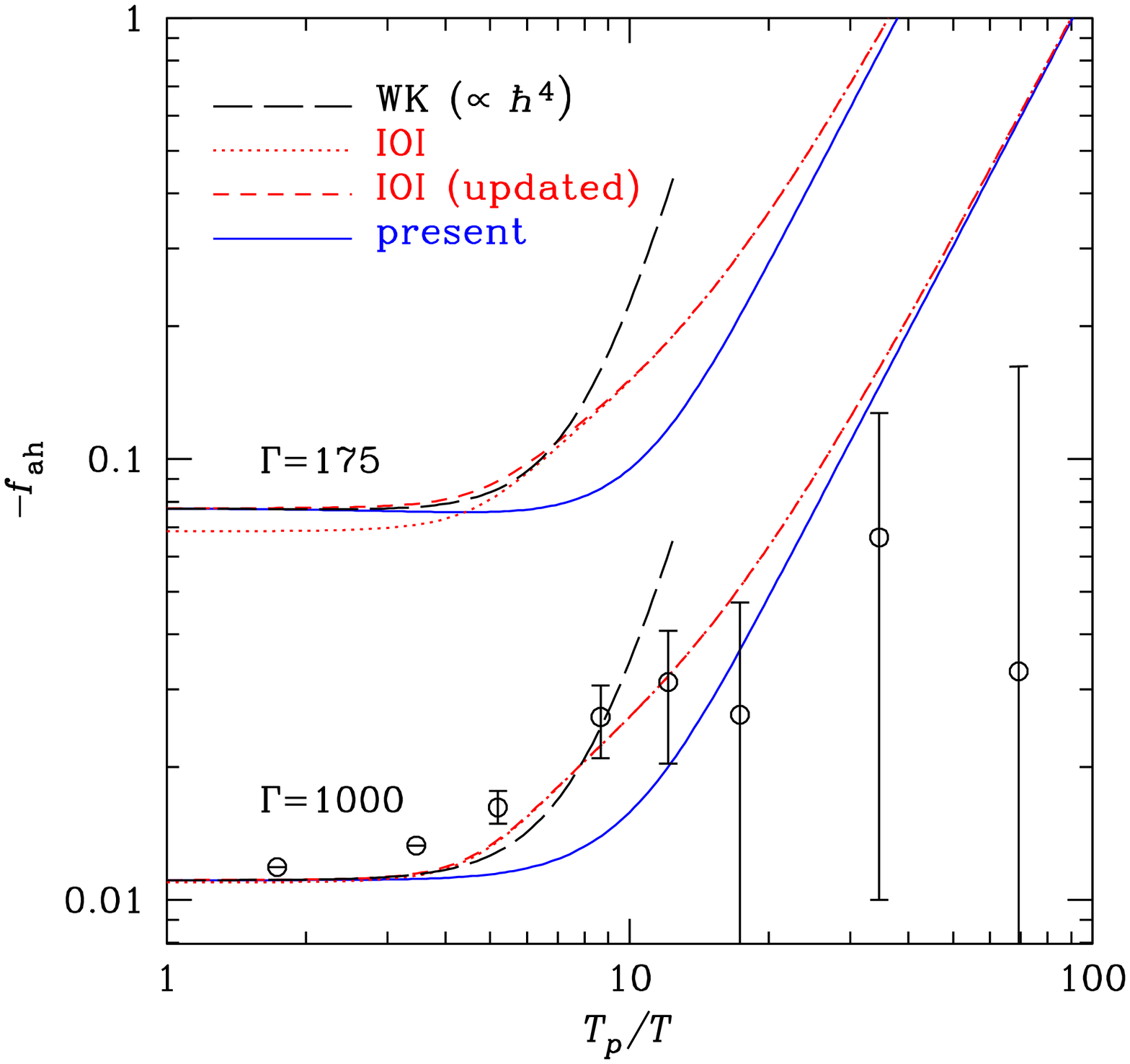}
\caption{Anharmonic contribution to the reduced free energy of an ion
lattice as a function of the quantum parameter $\eta=\Tp/T$ for
two values of the Coulomb coupling parameter, $\Gamma=175$ (upper
curves) and 1000 (lower curves).
A comparison of different approximations (see text):
Wigner expansion (long-dashed lines); an approximation
from \cite{Iyetomi-OI} (IOI, dotted lines); the same
approximation with the coefficients adjusted to
Ref.~\cite{FaroukiHamaguchi} (short-dashed lines); and present
interpolation (\ref{f_ah_fit}) (solid lines).
The points with errorbars show simulation results 
from \cite{Iyetomi-OI} for $\Gamma=1000$.
\label{fig:f_ah}
}
\end{minipage}
 \hfill
\begin{minipage}[t]{.48\linewidth}
    \includegraphics[width=\linewidth]{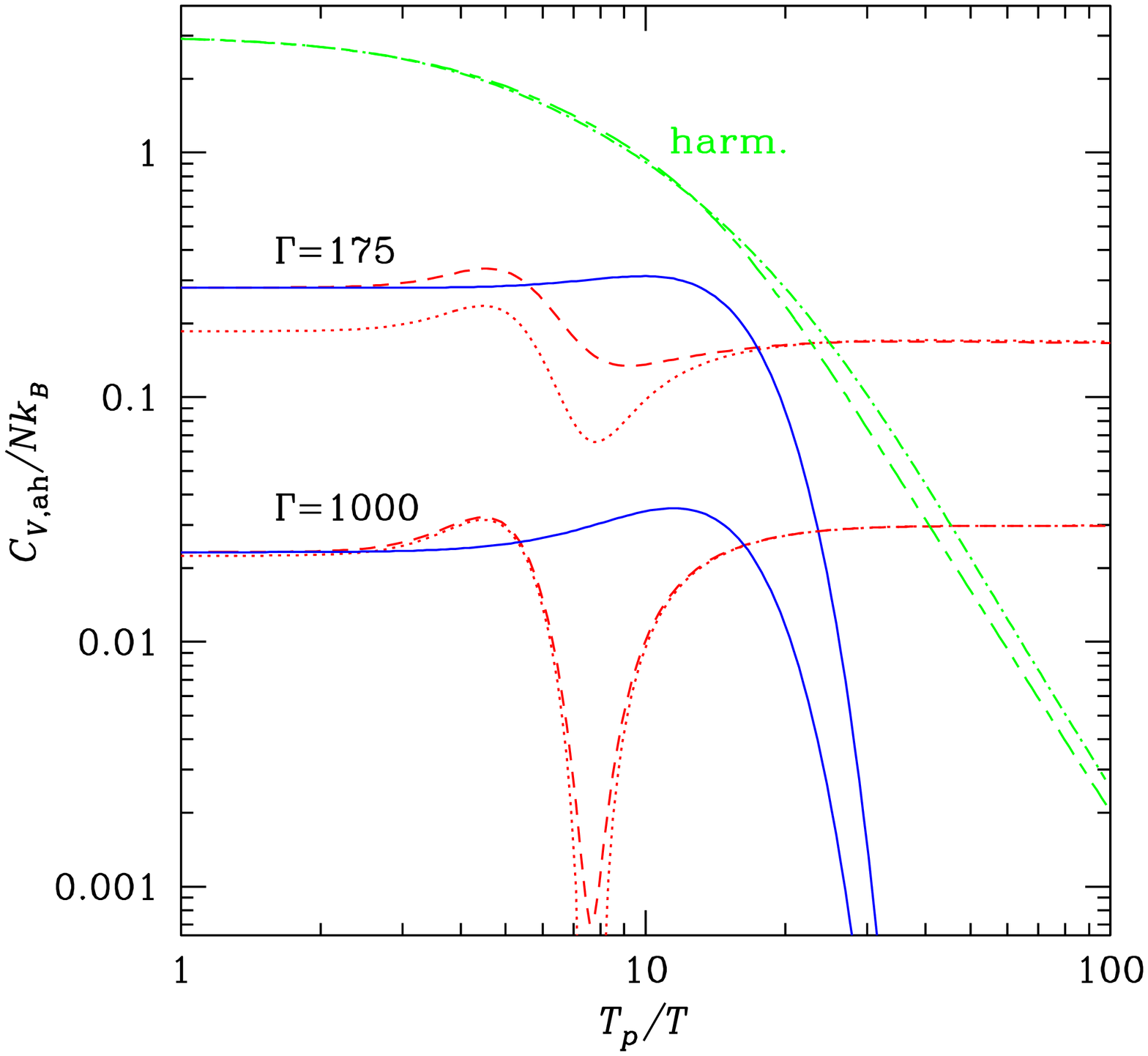}
\caption{Harmonic and anharmonic lattice contributions to the reduced heat
capacity at $\Gamma=175$ and 1000.
The harmonic lattice contribution according to
Ref.~\cite{Baiko-ea01} (dot-dashed lines) is compared to the
model \cite{Chabrier93} (long-short-dash lines) and to
the anharmonic correction in
different approximations (see text): 
a derivative of the IOI fit \cite{Iyetomi-OI} (dotted
lines: original; short-dashed lines: improved; see text) and
a corresponding derivative of
\req{f_ah_fit} (solid lines). 
}
\label{fig:cv_ah}
%\label{fig:f_ah}
\end{minipage}
\end{figure}

% ------------------------------
\subsection{Coulomb liquid}

For the reduced free energy of the ion-ion interaction
$f_\mathrm{ii}\equiv {F_\mathrm{ii}}/{\Nion\kB T}$
in the liquid OCP at any values of $\Gamma$ we use
the analytic approximation derived in Ref.~\cite{PC00}.
In order to extend its applicability range from $T\gg\Tp$
to $T\sim\Tp$, we add the lowest-order
quantum corrections to
the Helmholtz free energy
\cite{Hansen73}:
$
   f_q^{(2)}=\eta^2/24.
$
The next-order correction
$\propto\hbar^4$ has been obtained in \cite{HV75}.
These corrections have
limited applicability, because as soon as $\eta$ becomes large,
the Wigner expansion diverges and
the plasma forms a quantum liquid, whose free energy
is not known in an analytic form.

A classical OCP freezes at temperature $\Tm$ 
corresponding to $\Gamma=175$,
but in real plasmas $\Tm$ is affected by electron polarization
\cite{PC00,Hama-Yukawa} and quantum effects
(e.g., \cite{NNN,Chabrier93,JonesCeperley96}).
Hence the liquid does not freeze, regardless of $T$, at
densities larger than the critical one. 
The critical density values in the OCP correspond to
$R_S\approx 140$\,--\,160 for bosons and $R_S\approx 90$\,--\,110
for fermions. In astrophysical applications, the appearance of
quantum liquid can be important for hydrogen and helium
plasmas (see, e.g., Ref.~\cite{Chabrier93} for discussion).

% ------------------------------
\subsection{Coulomb crystal}

The reduced free energy of the Coulomb crystal is
$
   f_\mathrm{lat}\equiv
F_\mathrm{lat}/\Nion k_B T = C_0\Gamma + 1.5 u_1 \eta + f_{\mathrm{th}} +
   f_{\mathrm{ah}} .
$
Here, the first term is the Madelung energy ($C_0\approx-0.9$),
the second represents zero-point ion vibrations energy
($u_1\approx0.5$), $f_{\mathrm{th}}$ is the thermal correction in
the harmonic approximation,
 and $f_{\mathrm{ah}}$ is the anharmonic
correction. We use the most accurate values of $C_0$, $u_1$, and
analytic
approximations to $f_{\mathrm{th}}$ for bcc and fcc Coulomb OCP
lattices, which have been obtained in \cite{Baiko-ea01}.

For the \emph{classical anharmonic corrections}, 11
fitting expressions were given in \cite{FaroukiHamaguchi}. We have
chosen one of them:
$
   f_\mathrm{ah}^{(0)}(\Gamma) = a_1/\Gamma +a_2/2\Gamma^2 +
   a_3/3\Gamma^3,
$
with $a_1=10.9$, $a_2=247$, and $a_3=1.765\times10^5$, because
this choice is most consistent with the perturbation theory
\cite{NNN,Dubin}.

In applications one needs a continuous extension for the free
energy to $\eta\neq0$. 
With the leading quantum anharmonic correction at small $\eta$
one has \cite{HV75}
\beq
  f_\mathrm{ah} \approx f_\mathrm{ah}^{(0)}(\Gamma)
    - ( 0.0018/\Gamma + 0.085/\Gamma^2)\,\eta^4.
\label{ah-sc}
\eeq
At $T\to0$, the quantum anharmonic corrections 
were studied in \cite{Carr-ea,AlbersGubernatis,NNN},
where an expansion in powers of $R_S^{-1/2}$ was obtained,
assuming $R_S\gg1$. The leading
term of this expansion can be written in the form
$
   f_{\mathrm{ah},T\to0} = -b_1\eta^2/\Gamma.
$
In the literature one finds different
estimates for $b_1$; we use $b_1=0.12$ as an approximation
consistent with \cite{Carr-ea,AlbersGubernatis}. 
Free and internal energies of
\emph{finite-temperature} quantum crystals were calculated using
quantum Monte Carlo methods in 
\cite{Iyetomi-OI,JonesCeperley96,Chabrier-DP}.
Iyetomi et al.\ in Ref.~\cite{Iyetomi-OI} (hereafter IOI)
proposed
an analytic expression
for the quantum anharmonic corrections.
However, differences between the numerical results in
\cite{Iyetomi-OI,JonesCeperley96,Chabrier-DP}
are comparable to the differences
between the numerical results and the harmonic approximation. 
Therefore,
finite-temperature anharmonic corrections cannot be accurately determined
from the listed results.

In order to reproduce the zero-temperature and classical limits,
we multiply $f_\mathrm{ah}^{(0)}$ by exponential suppression factor
$e^{-c_1\eta^2}$ and add $f_{\mathrm{ah},T\to0}$.
Then we have
\beq
   f_\mathrm{ah} =  f_\mathrm{ah}^{(0)}(\Gamma)\,e^{-c_1\eta^2}
    -b_1\eta^2/\Gamma .
\label{f_ah_fit}
\eeq
According to \req{ah-sc},
the Taylor expansion coefficient at $\eta^2/\Gamma$ equals zero.
In order to reproduce this property, we set 
$c_1=b_1/a_1\approx0.0112$.
In Fig.~\ref{fig:f_ah}, the resulting approximation for the free
energy as a function of $\eta$ (solid lines) is compared
 to the
semiclassical expression [Eq.~(\ref{ah-sc}); long-dashed lines] and to
the elaborated IOI fitting formula:
the dotted lines correspond to the original coefficients
of this fit, chosen by IOI so as to reproduce
the results of Ref.~\cite{Dubin} at $\eta\to0$,
while the short-dashed curves
show the same fit, but with the coefficients adjusted to
more recent results \cite{FaroukiHamaguchi}
at $\eta\to0$. The curves of the two latter types nearly coincide at
$\Gamma=1000$, but differ near the melting point $\Gamma=175$.

Interpolation (\ref{f_ah_fit}) (unlike, for
example, Pad\'e approximations \cite{ChugunovBaiko}) does not produce an unphysical
behavior of thermodynamic functions. In particular, anharmonic
corrections to the heat capacity and entropy do not exceed
the reference (harmonic-lattice) values of these quantities
at any $\eta$.
In Fig.~\ref{fig:cv_ah}, we
show the ion contributions to the reduced heat capacity
of the ion lattice
$c_{V,i}\equiv C_{V,i}/\Nion\kB$, calculated through derivatives of
\req{f_ah_fit}
(solid lines), and compare it to the contributions calculated
through derivatives of the
fitting formula in \cite{Iyetomi-OI} (short-dashed and
dotted lines). In the same figure we plot
the harmonic-crystal contribution to the reduced heat capacity:
the short-dash--long-dash line corresponds to the model of
Ref.~\cite{Chabrier93} (which we adopted in \cite{PC00}),
while the dot-dashed line 
corresponds to the most accurate formula \cite{Baiko-ea01}.

Interpolation \req{f_ah_fit} should
be replaced by a more accurate formula in the future when
accurate finite-temperature anharmonic quantum corrections become
available.

%%%%%%%%%%%%%%%%%%%%%%%%%%%%%%%%%%%%%%%%%%%%%%%%%%%%%%%%%%%
\section{Electron polarization}
\label{sect:ie}
%%%%%%%%%%%%%%%%%%%%%%%%%%%%%%%%%%%%%%%%%%%%%%%%%%%%%%%%%%%

\subsection{Coulomb liquid}
\label{sect:ie-liq}

Electron polarization in Coulomb liquid was studied by
perturbation \cite{GalamHansen,YaSha} and HNC 
\cite{C90,ChabAsh,CP98,PC00}
techniques. The results for $F_{ie}$
are fitted by analytic expressions in \cite{PC00}.
This fit is accurate within several
percents, and it exactly
recovers the Debye-H\"uckel limit for EIP at $\Gamma\to0$ and
the Thomas-Fermi result
\cite{Salpeter61} at $\Gamma\gg1$ and $Z\gg1$.

%%%%%%%%%%%%%%%%%%%%%%%%%%%%%%%%%%%%%%%%%%%%%%%%%%%%%%%%%%%

\subsection{Coulomb crystal}
\label{sect:ie-sol}

Calculation of thermodynamic functions for a Coulomb crystal
with allowance for the electron polarization is a complex
problem. For classical ions, the simplest screening model
consists in replacing the Coulomb potential by the Yukawa
potential. For instance,
there were molecular dynamics simulations of classical
Yukawa systems 
\cite{Hama-Yukawa} and path-integral Monte Carlo simulations of
a quantum Yukawa crystal at $R_S=200$ \cite{MilitzerGraham}. 
However, the Yukawa interaction
reflects only the small-wavenumber asymptote of the electron
dielectric function.  A rigorous treatment would consist in
calculating the dynamical matrix and solving a corresponding
dispersion relation for the phonon spectrum. The first-order
perturbation approximation for the dynamical matrix of a
classical Coulomb solid with the polarization corrections was
developed in Ref.~\cite{PollockHansen}. The phonon spectrum in
such a quantum crystal has been calculated only in the harmonic
approximation \cite{Baiko02}, which has a restricted
applicability: for example, it cannot reproduce
the known classical ($T\gg \Tp$) limit of the
anharmonic $ie$ contribution
to the heat capacity.

A semiclassical perturbation approach was used in
\cite{PC00}. The results were fitted by
the analytic expression
\beq
   f_{ie} = -f_\infty(\xr)\,\Gamma
         \left[ 1 + A(\xr)\,(Q(\eta)/\Gamma)^s \right].
\label{fitscr-sol}
\eeq
In the classical Coulomb crystal,
$Q=1$,
$
f_\infty(x) = b_1\,\sqrt{1+b_2/x^2},
$
and
$
A(x) = (b_3+a_3 x^2 )/( 1+b_4 x^2)$.
Parameters $s$ and $b_1$--$b_4$ depend on $Z$ and are chosen
so as to fit the perturbational results; $a_3$ is a constant.
All the parameters weakly depend on the lattice type;
they are explicit in \cite{PC00} for the bcc and fcc lattices.

The factor $Q(\eta)$ in \req{fitscr-sol} is designed
to reproduce the suppression of the dependence of $F_{ie}$ on $T$
at $T\ll \Tp$. For the classical solid, $Q(0)=1$. 
In the quantum limit ($\eta\to\infty$)  $Q(\eta) \simeq
q\eta$, so that the ratio $Q/\Gamma$ in \req{fitscr-sol}
becomes independent of $T$. The 
proportionality coefficient $q$ was found numerically
in \cite{PC00}; it equals 0.205 for the bcc lattice.

The form of $Q(\eta)$ suggested in \cite{PC00}
assumed a too slow
decrease of the heat capacity ($C_{V,ie}\propto \eta^{-1}\propto
T$) at $\eta\to \infty$, which signals the violation of the
employed semiclassical perturbation theory in the strong quantum
limit, related to the approximate form of the ion
structure factor used in the calculations, as discussed in
\cite{PC00,NSB1}. In order to fix this problem, we
have changed the form of $Q(\eta)$ to
\beq
   Q(\eta) = \left[ { \ln\left(1+e^{(q\eta)^2} \right)
          }\right]^{1/2}\,\left[{
             \ln\left(e-(e-2)e^{-(q\eta)^2} \right)}
             \right]^{-1/2}.
\eeq
This form of $Q(\eta)$ has the correct limits at $\eta\to 0$ and $\eta\to
\infty$ and is compatible with the numerical
results \cite{PC00}. In addition, it eliminates the
problematic $ie$ contributions to the heat capacity and entropy
at $\eta\gg1$. It can be
improved in the future when the polarization corrections for the
quantum Coulomb crystal are accurately evaluated.

%%  SECTION   ---------------------------
\section{Conclusions}
\label{sect:concl}

We have reviewed analytic approximations for the
EOS of fully ionized electron-ion plasmas and described recent
improvements to the previously published approximations, taking
into account nonideality due to ion-ion, electron-electron, and
electron-ion interactions. The presented formulae are applicable
in a wide range of plasma parameters, including the domains of
nondegenerate and degenerate, nonrelativistic and relativistic
electrons, weakly and strongly coupled Coulomb liquids, classical
and quantum Coulomb crystals.

For brevity we have considered plasmas composed of electrons and
one type of ions. Extension to the case where several types of
ions are present is provided by 
\cite{PCR09,PCCDR09}.

We have made the Fortran code that realizes the analytical
approximations for the free energy and its derivatives, described
in this paper, freely available in the Internet
\footnote{http://www.ioffe.ru/astro/EIP/}.

\begin{acknowledgement}
The work of A.Y.P.\ was partially supported 
by the Rosnauka Grant NSh-2600.2008.2
and the RFBR Grant 08-02-00837,
and by CompStar, a Research Networking Programme of the
European Science Foundation. 
\end{acknowledgement}

% ----------------------------------------------------------------------

\end{document}